\documentclass[12pt]{article}
\usepackage{amsmath}
\usepackage{graphicx}
\usepackage{color}
\usepackage{enumerate}
\usepackage[authoryear,round,comma]{natbib}
\usepackage{url} 
\usepackage{hyperref}
\usepackage{caption}
\usepackage{subcaption}
\usepackage{setspace}
\usepackage{JASA_manu}

\usepackage{fancyhdr}
\newtheorem{assumption}{Assumption}
\newcommand{\qed}{\nobreak \ifvmode \relax \else
      \ifdim\lastskip<1.5em \hskip-\lastskip
      \hskip1.5em plus0em minus0.5em \fi \nobreak
      \vrule height0.75em width0.5em depth0.25em\fi}

\def\spacingset#1{\renewcommand{\baselinestretch}%
{#1}\small\normalsize} \spacingset{1}

\begin{document}

 \title{{\bf Nonparametric Bayesian Instrumental Variable Analysis: Evaluating Heterogeneous Effects of Coronary Arterial Access Site Strategies}}
  \author{Samrachana Adhikari$^1$, Sherri Rose$^2$, Sharon-Lise Normand$^{2,3}$ 
  \\
  1. Department of Population Health, New York University School of Medicine \\
  2. Department of Health Care Policy, Harvard Medical School\\
  3. Department of Biostatistics, T.H. Chan Harvard School of Public Health
   }
 \date{}
  \maketitle

\bigskip

\begin{abstract}
Percutaneous coronary interventions (PCIs) are nonsurgical procedures to open blocked blood vessels to the heart, frequently using a catheter to place a stent. The catheter can be inserted into the blood vessels using an artery in the groin or an artery in the wrist. Because clinical trials have indicated that access via the wrist may result in fewer post procedure complications, shortening the length of stay, and ultimately cost less than groin access, adoption of access via the wrist has been encouraged. However, patients treated in usual care are likely to differ from those participating in clinical trials, and there is reason to believe that the effectiveness of wrist access may differ between males and females. Moreover, the choice of artery access strategy is likely to be influenced by patient or physician unmeasured factors. To study the effectiveness of the two artery access site strategies on hospitalization charges, we use data from a state-mandated clinical registry including 7,963 patients undergoing PCI. A hierarchical Bayesian likelihood-based instrumental variable analysis under a latent index modeling framework is introduced to jointly model outcomes and treatment status. Our approach accounts for unobserved heterogeneity via a latent factor structure, and permits nonparametric error distributions with Dirichlet process mixture models. Our results demonstrate that artery access in the wrist reduces hospitalization charges compared to access in the groin, with a higher mean reduction for male patients.
\end{abstract}


\vfill

\spacingset{2} 

\section{Which Artery Access Site to Use?}
\label{sec::Introduction}

A percutaneous coronary intervention (PCI) is a nonsurgical procedure that uses a catheter (thin flexible tube) to place a small structure, usually a stent, in the blocked blood vessels that have been narrowed by plaque buildup. The catheter could be inserted into the blood vessels either via an artery in the groin (a femoral artery), or via an artery in the wrist (a radial artery). There are over one million PCIs performed each year in the United States, predominantly using femoral arterial access \citep{feldman2013adoption, Bader2017}. Early complications after a PCI include bleeding from the arterial access site as well as vascular complications such as blood transfusions, hemorrhagic stroke, and cerebrovascular accidents. Randomized controlled trials (RCTs) have indicated that these complications occur less frequently after radial arterial access compared to femoral arterial access PCI \citep{ferrante2016radial}, and these fewer complications may likely lead to lower hospitalization charges \citep{RadialFemoralCost2013, AMIN2013827}. However, such differences in hospitalization charges and in complications between the arterial access sites have not been well-studied in an observational setting where patients and physicians are likely to differ from those in RCTs. 

There are many methodological challenges when comparing the effectiveness of radial and femoral arterial access strategies in observational data. First, healthier patients are often selected to undergo radial arterial access PCI, thus introducing selection bias \citep{RAO2008379}. It is, however, almost impossible to have complete measurements for all the health indicators that could have contributed to a physician's decision to perform a radial rather than a femoral arterial access PCI. Second, RCTs that have studied the effectiveness of the two access strategies also reported evidence of treatment effect heterogeneity \citep{RAO2008379}. For instance, female patients tend to have fewer bleeding and vascular complications with radial arterial access versus femoral arterial access, compared to male patients. With the recent push to encourage adoption of radial arterial access PCI \citep{feldman2013adoption}, it is important to understand the comparative effectiveness of the two arterial access site strategies in observational settings. To do so, we make use of a state-mandated clinical registry coordinated by the Massachusetts Data Analysis Center (Mass-DAC)  for 7,963 adults undergoing PCI \citep{doi:10.1056/NEJMoa0801485}. Our goal is to estimate heterogeneous access site effects based on sex of patients in observational data where there is also selection bias.

To compare the effectiveness of radial and femoral arterial access strategies in observational data, we investigate likelihood-based estimators for instrumental variable analysis using a latent index model framework \citep{heckman1999local, ECTA:ECTA277}. Under this framework, model assumptions are jointly specified on treatment and potential outcomes. Necessary assumptions along with sensitivity analysis for the assumptions to make a valid causal inference based on such estimators are assessed. Our focus is on a Bayesian latent factor approach, which is a nonparametric extension of the approach proposed by \citet{heckman2014treatment}, to make posterior inference on the estimates of causal effects, with particular emphasis in settings where there is treatment effect heterogeneity. Our approach has several practical advantages over earlier approaches. It provides a framework to model complex latent structures and account for non homogeneous correlation structures that standard parametric approaches do not. We do not rely on parametric distribution of errors. Further, Bayesian methodology is used to estimate and make posterior inference on causal parameters for which analytical estimates can be intractable. Simulations are developed to characterize operating characteristics of various estimators. Finally, we implement our approach to study heterogeneous treatment effects of radial arterial access compared to femoral arterial access PCI on hospitalization charges based on sex of the patients. 

The organization of the paper is as follows. We introduce the data and provide background literature for latent index models in Section \ref{sec::background}. We then introduce our proposed approach using a nonparametric Bayesian model in the context of latent index modeling with latent factors, along with parameter estimation and derivation of treatment effects in Section \ref{sec::IV_DPM}. We evaluate the performance of the proposed approach via simulation studies in Section \ref{sec::simulation} and illustrate our strategy using the PCI data in Section \ref{sec::dataintro}. We end with discussion in Section \ref{sec::conclusion}.

\section{Background}
\label{sec::background}

\subsection{Data and Outcomes}
\label{ssec::application}
The state-mandated clinical registry coordinated by the Mass-DAC \citep{doi:10.1056/NEJMoa0801485} includes all PCIs performed in all nonfederal acute care Massachusetts hospitals for patients at least 18 years of age, regardless of health insurance status. The registry data are populated by trained data managers at each hospital.  Baseline covariates include demographic information, such as age, sex, race, types of health insurance; comorbidities and  family history of cardiac problems; cardiac presentation prior to the PCI, such as ejection fraction, cardiac shock, acute coronary syndrome status; procedure-specific information, such as number of treated vessels, degree of blockage, artery access strategy, type of stent inserted; in-hospital complications; discharge medications; and survival status. The registry data are linked to the Massachusetts Acute Hospital Case-Mix billing data maintained by the Massachusetts Center for Health Information and Analysis to obtain additional post-discharge hospitalization information, such as subsequent admissions for PCI, bypass surgery, heart attack, etc. Information is linked using criteria based on combinations of treatment hospital, medical record number, admission or discharge date, and date of birth. Vital status is determined using the Massachusetts Registry of Vital Statistics as well as the National Death Index. 

We focus on 7,963 adults $\ge 18$ years old undergoing PCI in 2011. Patients who resided outside Massachusetts (to ensure completeness of 30-day follow-up) and those who were deemed to be of exceptional risk, defined as patients having high-risk features not captured by any variable in the data or cases where PCI offered the best or only option for improving the chance of survival, are excluded. Because patients can undergo more than one PCI during a hospitalization, the first or index PCI during the hospitalization is analyzed (see Supplemental Material for a complete list of covariates).

\begin{figure}
\centering
\includegraphics[height=3in,width=3.5in]{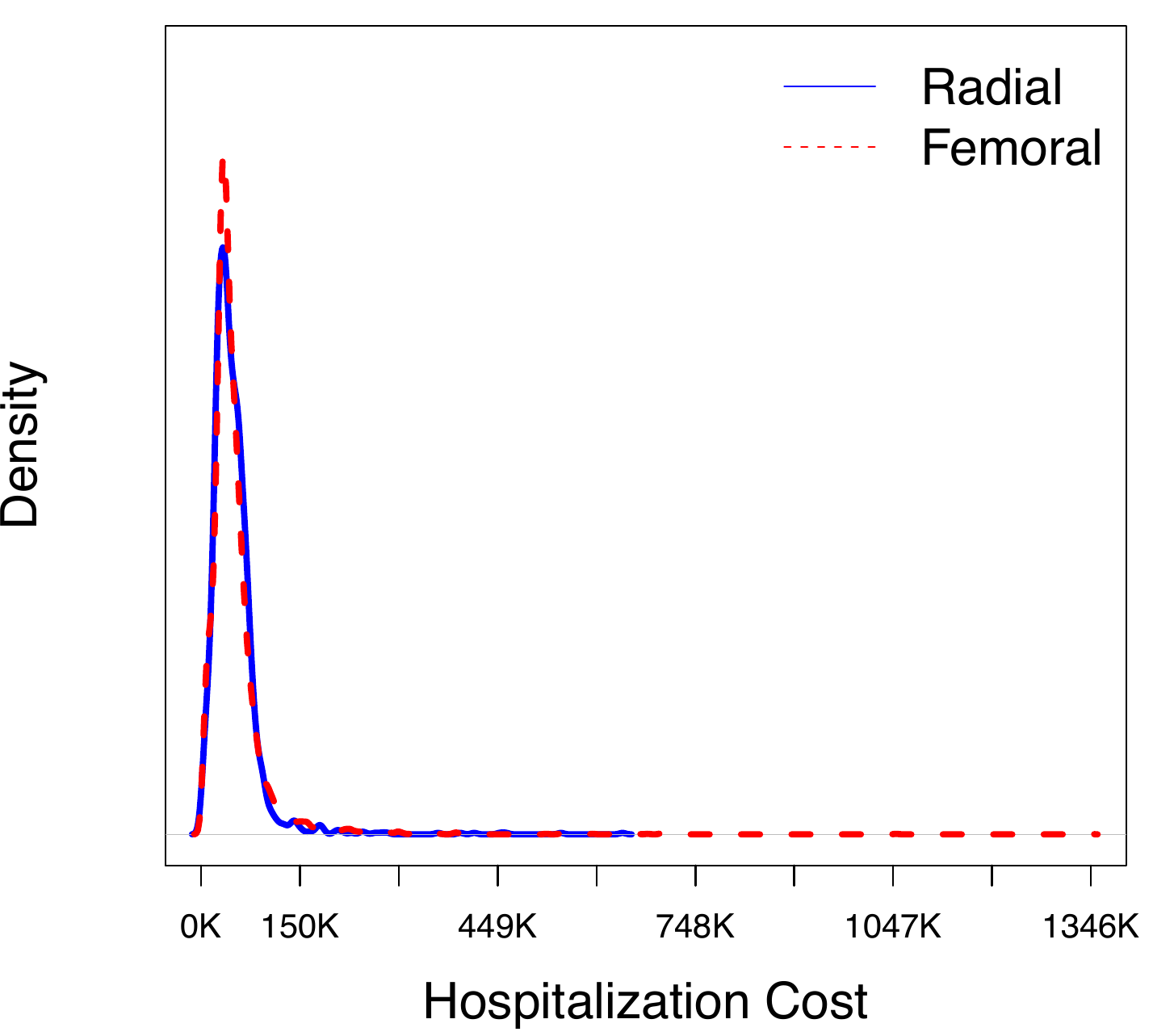}
\caption{\bf Density of observed hospitalization charges for radial and femoral PCI separately.}
\label{fig::costByTreatment}
\end{figure}

The main endpoint of interest is the health care (procedure-related) charges at hospital discharge. This is defined as the sum of full, undiscounted charges for patient care summarized by prescribed revenue codes for special care, routine accommodation, and ancillary services. Total charges do not include charges for telephone service, television, or private duty nurses. As expected, the distributions of total charges at discharge for patients with radial and femoral arterial access PCIs are skewed (Figure \ref{fig::costByTreatment}).

Additional outcomes, including 30-day and 1-year hospitalizations for major bleeding and in-hospital vascular complications, and  a falsification outcome, target vessel revascularization (TVR), are also considered. TVR is defined as PCI performed in a vessel treated during the index procedure or any coronary artery bypass graft (CABG) surgery performed within 1-year of undergoing the index PCI. TVR is a falsification outcome because no change in TVR is expected as a consequence of different arterial access \citep{mauri2008long}. 

Table \ref{table::OutcomebyTrGe} summarizes outcomes for patients with radial arterial access and femoral arterial access PCI stratified by sex. Of the 7,963 patients who had PCI in Massachusetts during 2011, the majority (69\%) were male and only 23\% involved radial arterial access PCI. Patients who had radial arterial access PCI had overall lower mean hospitalization charges with lower incidences of bleeding and vascular complications compared to those with femoral arterial access. We observe a two percentage point difference in the observed incidences of TVR between patients who had radial arterial access PCI versus femoral arterial access PCI.
 \begin{table}[!ht]
 \centering
 \begin{tabular}{rrrrrrr}
   \hline
  & \multicolumn{2}{c}{\bf Males}  &  \multicolumn{2}{c}{\bf Females} & \multicolumn{2}{c}{\bf Overall } \\ \cline{2-7}
{\bf Arterial Access Site} & Radial & Femoral & Radial & Femoral & Radial & Femoral  \\ \hline
No. PCI Admissions &  1,325 & 4,141 &  498 & 1,999&	1,823 & 6,140 \\  \hline
{\bf In-Hospital Outcome} &&&&&&\\
   Mean Total Charges, \$ & 48,718 & 53,539 & 49,274 & 51,107 & 48,878 & 52,747 \\ 
   Vascular Complication,\% & 1.4 & 4.0 & 2.8 & 8.4 & 1.7 & 5.4 \\ \hline
   30-day Major Bleed, \% & 1.6 & 4.0 & 2.6 & 7.2  & 1.8 & 5.0 \\ \hline
{\bf 1-Year Outcomes, \%} &&&&&&\\
Major Bleed & 7.2 & 8.2 & 11.0 & 17.2  & 8.2 & 13.3\\ 
TVR & 6.0 & 8.2 & 6.6 & 8.2 & 6.2 & 8.2 \\ 
\hline
 \end{tabular}
\caption{{\bf Unadjusted mean outcomes by treatment \& sex.} Bleed defined by an in-hospital admission for major bleeding.}
\label{table::OutcomebyTrGe}
 \end{table}
 
The goal of this study is to estimate heterogeneous effects of the two arterial access site strategies on hospitalization charges, by adjusting for selection bias due to unmeasured confounding, using instrumental variable analysis. Analysis of the PCI data is described in detail in Section \ref{sec::dataintro}. An instrument for the analysis can be defined based on contextual knowledge. Validity of the instrument is assessed with sensitivity analysis using the outcomes and baseline covariates described in this section, before proceeding to estimate treatment effects.

\subsection{Approaches to Inference in the Presence of Unmeasured Confounders}
\begin{figure}
\centering
\includegraphics[scale=0.6]{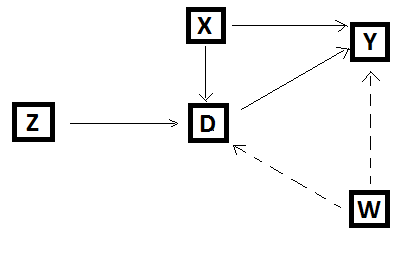}
\caption{{\bf A directed causal graph for IV analysis.} Arrow display the relationships among the (potential) outcomes $Y$, treatment $D$, measured confounders $X$, instrument $Z$ and unmeasured confounders $W$.}
\label{fig::IVdag}
\end{figure}

An instrumental variable (IV) analysis approach permits causal inference with a confounded treatment assignment mechanism. An IV, or simply an instrument, is defined as a random variable that can predict treatment, but is independent of potential outcomes conditional on observed covariates \citep{angrist1996identification}. This relationship among the (potential) outcomes $Y$, treatment $D$, measured confounders $X$, instrument $Z$, and unmeasured confounders $W$ is depicted in Figure \ref{fig::IVdag}. 
Likelihood-based estimators of treatment effects in IV analysis have been introduced in previous literature. \citet{chib2000bayesian,carneiro2003understanding}, and \citet{o2011estimating} proposed bivariate probit (or joint Normal) models to simultaneously model treatment and outcomes to estimate mean treatment effect. \citet{heckman2014treatment} and \citet{jacobi2016bayesian} investigated Bayesian latent factor models to go beyond mean treatment effects and estimate distributional effects. However, the model assumptions discussed in \citet{heckman2014treatment} and \citet{jacobi2016bayesian} may be too restrictive if, for example, the continuous outcome has a non-Normal distribution. This is the case in our application, where densities of observed charges for the two different arterial access strategies are heavily skewed. 

An extensive literature exists on alternative methods for estimating causal effects using IV analysis that utilize a method of moments approach within a marginal structural model framework \citep{hernan2004structural, van2007estimation, tan2010marginal}, including the two-stage least squares estimation approach \citep{angrist1996identification, SIM:SIM6128}. While the method of moments approach does not make any distributional assumptions, inference often relies on limiting assumptions, such as constant treatment effect, for identification of causal effects. Heterogeneous treatment effects in an IV setting is a recent field of investigation with limited work \citep{HEC:HEC1291, tan2010marginal, heckman2014treatment}. Our paper adds to this burgeoning literature.

\subsection{Latent Index Model for IV Analysis}
The causal effects are defined using the potential outcome framework of \citet{rubin1974estimating}. Let $D_i \in \{0,1\}$ denote a binary treatment applied to patient $i$ and let $(Y_i(0),Y_i(1))$ denote the two corresponding continuous potential outcomes of patient $i$ under each treatment status. The stable unit treatment value assumption (SUTVA) \citep{rubin1990formal, angrist1996identification} is assumed such that (a) the treatment assignment of one patient does not interfere with the outcome of another patient, and (b) radial arterial access strategy is well-defined and the same for all patients who receive it and femoral artery access strategy is well-defined and the same for all patients who receive it. 

A latent index model for IV analysis can then be written as
\begin{equation}\label{eq::heckmanIV}
\begin{aligned}
D &= \mathbf{I}(D^{*} > 0) ~~\mbox{where}  ~~D^{*} = f(Z,X) + U_D \\ 
Y(0) &= g_0(X) + U_0 ~~\mbox{and}~~ Y(1) = g_1(X) + U_1.
\end{aligned}
\end{equation}
Here, $D^{*}$ is a latent continuous treatment variable, $\mathbf{I}(\cdot)$ is an indicator function such that $D$ evaluates to one if condition $D^{*}>0$ is satisfied and to zero otherwise; $Z$ is used to denote instrumental variable(s); $X$ denotes a matrix of observed confounders; $f$ is a function relating the instruments and the confounders to the latent treatment status $D^{*}$; $g_0$ and $g_1$ are functions that relate confounders with potential outcomes; and $U_D$, $U_0$ and $U_1$ are corresponding errors in the model. While $g_0$, $g_1$ and $f$ can be nonlinear, semiparametric, or nonparametric functions \citep{chib2007semiparametric}, most of the existing work makes linear, parametric assumptions \citep{chib2000bayesian,heckman2014treatment}.  It is further assumed that $X$ and $Z$ are fixed. 

When there are unobserved confounders, treatment assignment is not ignorable, i.e., $U_0$ and $U_1$ are not independent of $D$ given $X$. Validity of the IV analysis to identify a causal effect, in such cases, is based on whether an instrument $Z$ meets ``exclusion restriction" assumptions \citep{angrist1996identification} stated in Assumption \ref{ass:er} below.
\begin{assumption}{}
\label{ass:er}
A) Probability of a treatment choice is a nontrivial function of the instrument $Z$ conditional on covariates $X$. 
B) Instrument $Z$ is mean independent of the error terms $U_0$ and $U_1$ in Equation \ref{eq::heckmanIV}, conditional on $X$.
\end{assumption}
The implication of Assumption \ref{ass:er} is that there is no direct effect of the instrument on the outcome. The instrument $Z$ affects the potential outcomes $(Y(1),Y(0))$ only through its effect on the treatment $D$. 

In the latent index model (Equation \ref{eq::heckmanIV}), there is also an implicit assumption of monotonicity \citep{angrist1996identification, ECTA:ECTA277, tan2006regression} as stated in Assumption \ref{ass:mono} below.
\begin{assumption}\label{ass:mono}
Let $Z=Z(\omega)$ be a vector of instruments and $D_z = D_z(\omega)$ be the treatment status that would be observed if $Z(\omega)$ were externally set to $z$. For any two levels of instrument $z$ and $z^{'}$ and vector of confounders $x$, monotonicity is defined as either $D_z(\omega) \geq D_z^{'}(\omega)$ for all $\omega \in \{X=x\}$ or  $D_z(\omega) \leq D_z^{'}(\omega)$ for all $\omega \in \{X=x\}$.
\end{assumption}
Assumptions \ref{ass:er} and \ref{ass:mono} are required to estimate average treatment effects under various conditions in a frequentist setting \citep{abadie2002bootstrap, HEC:HEC1291}. Earlier work on Bayesian methodologies for estimating average treatment effects of interventions with selection bias have assumed Normal error distribution for potential outcome models, and very few are concerned with estimating heterogeneous treatment effects \citep{chib2000bayesian, hirano2000assessing,heckman2014treatment, jacobi2016bayesian, choi2017estimating}.

The variance-covariance matrix of the error terms in Equation \ref{eq::heckmanIV} is given by
$$\textrm{COV}\begin{pmatrix}U_D \\ U_1 \\ U_0 \end{pmatrix} =
\begin{pmatrix} \sigma_D^2 & \rho_{D1}\sigma_1 & \rho_{D0}\sigma_0 \\
                \rho_{D1}\sigma_1 & \sigma_1^2 & \rho_{10}\sigma_1\sigma_0 \\
                \rho_{D0}\sigma_0 & \rho_{10}\sigma_1\sigma_0 & \sigma_0^2           
                     \end{pmatrix},$$
where, $\sigma^2_D$ is the variance in treatment equation, $\sigma^2_1$ and $\sigma^2_0$ are the variances of the potential outcomes $Y(1)$ and $Y(0)$ respectively, and $\rho_{D1}$ is the covariance between the treatment $D$ and $Y(1)$, and so on. Thus, the latent index model in Equation \ref{eq::heckmanIV} also assumes that the variance-covariance matrix is homogeneous for everyone in the sample. This assumption of homogeneous variance can be unrealistic when the underlying treatment selection might vary by different patient subgroups, such as patients treated in different locations and hospitals. Additionally, if the parameter of interest is an average treatment effect, the covariance parameter between the potential outcomes, $\rho_{10}$, is usually assumed to be zero \citep{chib2000bayesian, o2011estimating,  choi2017estimating}. However, for estimating effects that are dependent on estimating distribution of $Y_i(1) - Y_i(0)$, or some functionals of it, such as identifying fractions of a population that benefit from a given treatment, these approaches are inadequate.  The correlation between $Y(1)$ and $Y(0)$ needs to be accounted for \citep{heckman2014treatment}, but because $Y_i(1)$ and $Y_i(0)$ are not observed for the same patient $i$, additional assumptions are necessary to identify $\rho_{10}$. A latent factor structure is utilized in Section \ref{sec::IV_DPM} to address these shortcomings.

\section{Nonparametric Residual Bayesian Latent Factors Model}
\label{sec::IV_DPM}

To account for heterogeneous variance in the selection process of the latent index model presented in Equation \ref{eq::heckmanIV}, we extend the work of \cite{heckman2014treatment} on Bayesian models for continuous potential outcomes with a latent factor structure, and introduce a nonparametric residual Bayesian latent factors model. The latent index model with a latent factor structure can be written as
\begin{equation}\label{eq::heckmanIV_Latent}
\begin{aligned}
D &= \textbf{I} (D^{*} > 0) ~~\mbox{where}~~ D^{*} = \beta_D^0 +\gamma  Z + \beta_DX + \alpha_D \Theta+\epsilon_D \\
Y(1) &= \beta_1^0 + \beta_1X + \alpha_1 \Theta + \epsilon_1 ~~\mbox{and}~~ Y(0) = \beta_0^0 + \beta_0X + \alpha_0 \Theta + \epsilon_0. 
\end{aligned}
\end{equation}
The parameters $\gamma$, $\beta_D$, $\beta_1$ and $\beta_0$ are regression coefficients; $\Theta$ is a matrix of latent factors that explain the unobserved heterogeneity in the potential outcomes and the treatment selection models, such that $Y(1),Y(0)$ and $D$ are assumed to be independent conditional on $\Theta$; and $\alpha_D$, $\alpha_1$ and $\alpha_0$ are corresponding factor loadings for the treatment model and potential outcomes models, respectively.

All the dependence between the unobservables in the model is driven by the latent factor $\Theta$. By restricting $\Theta$ to be of low dimension, the covariance matrix of the error term is fully identified by the latent structure of the model \citep{heckman2014treatment}. We can use the latent representation to understand the correlation structure between potential outcomes and treatment assignments in the proposed model. First note that
\begin{eqnarray*}
\textrm{COV}(pr(D=1),Y(1))   
& =& \textrm{COV}(\beta_D^0 +\gamma  Z + \beta_DX + \alpha_D \Theta+\epsilon_D,\beta_1^0 + \beta_1X + \alpha_1 \Theta + \epsilon_1 ) \\
 & = & \alpha_D\alpha_1\textrm{Variance}(\Theta), 
\end{eqnarray*}
where the last equality follows because $Z$ is independent of $Y(1)$ from Assumption \ref{ass:er} (exclusion restriction) and the assumption that $\epsilon_D$ and $\epsilon_1$ are independent.  Similarly, it can be shown that $\textrm{COV}(pr(D=1),Y(0)) = \alpha_D\alpha_0\textrm{Variance}(\Theta)$, and $\textrm{COV}(Y(1),Y(0)) = \alpha_1\alpha_0\textrm{Variance}(\Theta)$;
$\textrm{Variance}(U_D)  = \alpha_D^T\textrm{Variance}(\Theta)\alpha_D + 1,$
$\textrm{Variance}(U_1) =  \alpha_D^T\textrm{Variance}(\Theta)\alpha_D + \sigma_1^2,$ and
$\textrm{Variance}(U_0) =   \alpha_D^T\textrm{Variance}(\Theta)\alpha_D + \sigma_0^2.$

The nonparametric residual assumption relies on utilizing Dirichlet process mixture (DPM) priors \citep{ferguson1973bayesian} on $\epsilon_1$ and $\epsilon_0$, to build a new flexible method for understanding the heterogeneity and permitting dependence between the treatment $D$ and the continuous potential outcomes $Y(1)$ and $Y(0)$. We refer to our proposed extension as the DPM latent index variable (LIV) model. The DPM prior on the errors, $\epsilon_d$ for $d \in \{0,1\}$, of potential outcome under each treatment status $d$ for patient $i$ in the DPM LIV model can be written as
\begin{equation}
\label{eq::IV_DPM}
\begin{aligned}
\epsilon_{d,i} & \sim  \textrm{Normal}(\mu_{d,i},\sigma^2_{d,i}) \\
(\mu_{d,i},\sigma^2_{d,i})|\mathcal{P} &\sim \mathcal{P} = \textrm{Dirichlet Process}(c^d,G_0) \\
 \equiv \mathcal{P} &\sim \mathcal{P}_{\infty} \textrm{ s.t. } \mathcal{P}_{\infty}(.) = \sum_{j=1}^\infty \mathcal{P}_j\delta_j(.),
\end{aligned}
\end{equation}
where, $c^d$ is a concentration parameter for treatment status $d$, $G_0$ is a base distribution and $\delta_j(.)$ is a Dirac delta function. The last line in Equation \ref{eq::IV_DPM} suggests that a Dirichlet process can also be seen as an infinite-dimensional generalization of Dirichlet distribution and is a conjugate prior for infinite, nonparametric discrete distributions. We use conjugate Normal($\omega$, $\tau$) as a base distribution $G_0$. Further prior specifications for the DPM process are described in Subsection \ref{ssec::estimation}. It is assumed that the errors $\epsilon_D$, $\epsilon_1$ and $\epsilon_0$ are jointly independent, and $\textrm{Variance}(\epsilon_D) = 1$, $\textrm{Variance}(\epsilon_1) = \sigma_1^2 < \infty$, $\textrm{Variance}(\epsilon_0) = \sigma_0^2 < \infty$ . 

\cite{heckman2014treatment} assume the errors $\epsilon_1$ and $\epsilon_0$ in the potential outcome models have symmetric Normal distributions. This specification can further introduce bias and estimation uncertainty if the underlying densities of $Y(1)$ and $Y(0)$ are more complicated with many modes or thick tails \citep{marra2016simultaneous}. Our implementation of a DPM LIV model makes a less restrictive assumption on the distribution of error and addresses the issues of bias and estimation uncertainty that arise from misspecification of the error distribution.


\subsection{Estimation of Model Parameters}
\label{ssec::estimation}
Metropolis Hastings within a Gibbs Markov Chain Monte Carlo (MCMC) algorithm is used to sample from the posterior distribution of the parameters given observed data. The joint posterior distribution of the model parameters, $\Omega$, conditional on observed data can be written as a product of likelihood and prior, $ P(Y_{obs},D_{obs}|X,Z,\Omega)\times \Pi (\Omega)$. The likelihood part of the product in the joint posterior can be factorized as 
\begin{equation}
\label{eq::lik}
\begin{aligned}
& P(Y_{obs}|D_{obs},X,Z,\Omega) P(D_{obs}|X,Z,\Omega)\\
&= \prod_{i \in \{D_i=1\}} P(Y_{i}|D_i,\beta_1^0,\alpha_1,\beta_1,\Theta_i,X_i,Z_i,\mu_{1,i},\sigma_{1,i}^2) \\
&  \ \ \ \  \prod_{i \in \{D_i=0\}} P(Y_{i}|D_i,\beta_0^0,\alpha_0,\beta_0,\Theta_i,X_i,Z_i,\mu_{0,i},\sigma_{0,i}^2)
 \prod_{i = 1}^n P(D_i=1|\beta_D^0,\gamma,\alpha_D, \beta_D,\Theta,X_i,Z_i) \\
&= \prod_{i \in \{D_i=1\}} P(Y_{i}(1)|\beta_1^0,\alpha_1,\beta_1,\Theta_i,X_i,\mu_{1,i},\sigma_{1,i}^2) \\
&\ \ \ \   \prod_{i \in \{D_i=0\}} P(Y_{i}(0)|\beta_0^0,\alpha_0,\beta_0,\Theta,X_i,\mu_{0,i},\sigma_{0,i}^2) 
 \prod_{i = 1}^n P(D_i=1|\beta_D^0,\gamma,\alpha_D, \beta_D,\Theta,X_i,Z_i), 
\end{aligned}
\end{equation}
where, $\mu_1$ and  $\sigma_1^2$ are vectors of means and variances on the errors of $Y(1)$, and $\mu_0$, $\sigma_0^2$ are vectors of means and variances on the errors of $Y(0)$ respectively. 
A combination of conditional independence and exclusion restriction assumptions is used to express the likelihood of observed data in terms of potential outcomes in Equation \ref{eq::lik}. For computational convenience and (statistical) identification we assume $E(\epsilon_0) = E(\epsilon_1) = E(\epsilon_D) = 0$. This assumption is equivalent to fitting the latent index model without an intercept and assuming DPM priors on the parameters of the error distribution for potential outcomes. Because we we have binary treatments, without loss of generality, we can fix the variance of $U_D$ (or equivalently $\sigma_D^2$) to 1.

We assume a Normal(0, 1) prior on $\Theta$, and non-informative Normal priors on the coefficients in the outcome models and treatment models. Additionally, the Inverse-Wishart$(\Psi, \nu)$ hyper-prior is used on the variance $\tau$ of the DPM prior such that $E(\tau) = \frac{\Psi^{-1}}{(\nu-1})$ and Normal-Inverse Gamma hyper-prior on the mean $\omega$ of the DPM prior. A Gamma(a,b) prior is employed on the concentration parameter $c$. The latent factor $\Theta$, the parameters of the outcome models, and the parameters of the treatment model are sequentially updated using Gibbs sampling. The parameters of the DPM model priors are updated using the stick breaking model of \citet{ishwaran2001gibbs}. The \textbf{R} package \texttt{DPpackage} \citep{jara2011dppackage} is implemented within our sampler to sample from the Dirichlet process mixture model and hence our prior representation for DPM prior is consistent with the convention in the package. The \textbf{R}-code for implementing the MCMC algorithm is publicly available in GitHub repositories, as an R-package in \url{https://github.com/SamAdhikari/BayesIV_0.1} and corresponding simulation study in \url{https://github.com/SamAdhikari/BayesIV_Simulations}.

It is useful to observe that $\Theta$ enters the likelihood of the potential outcomes via both the mean and the variance, such that 
$E(Y_i(0)) = E(X_i\beta_1+\alpha_0\Theta_i+\mu_{0,i})$ and $\textrm{Variance}(Y_i(0)) = 
\textrm{Variance}(\alpha_0\Theta_i)+\sigma_{0,i}^2$. We can derive similar formulae for $E(Y_i(1))$ and $\textrm{Variance}(Y_i(1))$. Thus, the $\alpha$s and $\Theta$ can only be identified jointly as a product. For identification, the prior variance of $\Theta$ is constrained by fixing the variance of Normal distribution at 1. Another identification issue arises in estimating the signs of the latent factor $\Theta$ and the factor loading $\alpha_k$, for $k \in \{0,1,D\}$, because the likelihood in Equation \ref{eq::lik} only depends on the sign of the product $\alpha_k\Theta$. The nonidentifiability of the sign is accounted for in the sampling scheme by random sign switch. 

\subsection{Treatment Effects in Bayesian DPM LIV Model}
\label{sec::effects}
We define different sample version of causal effects by averaging the outcome gain, $\Delta_i \equiv Y_i(1) - Y_i(0)$, below. The {\textit{average treatment effect (ATE)}} for a given set of covariates is defined as, 
$\textrm{ATE}(x) = E(\Delta|X=x) = \int \Delta p(\Delta|X=x)d\Delta.$ The covariate specific heterogeneity, which is of interest in our application, can be studied by summarizing $E(Y(1)-Y(0)|X_l=x)$ for groups at specific levels of a covariate (or covariates) $X_l$ of interest, and is referred to as the {\textit{conditional average treatment effect (CATE)}}. Finally, the {\textit{average treatment effect on the treated (ATT)}} for a given set of covariates can be defined as,
{$\textrm{ATT}(z,x) = E(\Delta|D=1,Z=z,X=x) = \int \Delta p(\Delta|D=1,Z=z,X=x)d\Delta.$} Our Bayesian approach also permits estimation of the {\textit{probability of benefitting (PB)}} from the treatment conditional on the covariates as, $\textrm{PB}(x)  = Pr(Y(1) > Y(0) |X)  = 1 - Pr(Y(1)-Y(0) < 0)$, using the posterior chains from MCMC. If instead, the interest is in evaluating whether the difference in the outcomes is greater than certain threshold, we can redefine PB$(x)$ by replacing 0 with the threshold. When there are heterogeneous treatment effects, the ATE(x), CATE(x), and ATT(x) will differ.

\subsection{Posterior Inference}
MCMC samples from the posterior distribution of the parameters conditional on observed outcomes, treatment, covariates and instruments can be used to obtain posterior estimates of the causal parameters defined below. First note that the joint distribution of the treatment and the outcomes conditional on covariate, instrument, latent factor $\Theta$ and set of model parameters $\Omega$ is
$$\begin{pmatrix} 
D_i^{*}\\Y_i(1) \\Y_i(0)
\end{pmatrix}
|\Theta_i,X_i,Z_i, \Omega
\sim \textrm{Normal}
\left(\begin{pmatrix}\gamma Z_i + \beta_D X_i + \alpha_D \Theta_i \\ 
\beta_1X_i + \alpha_1\Theta_i + E(\epsilon_{1,i}) \\
 \beta_0 X_i+\alpha_0\Theta_i + E(\epsilon_{0,i})  
 \end{pmatrix}, 
 \begin{pmatrix}1 & 0 & 0 \\0 & \textrm{Var}(\epsilon_{1,i}) & 0\\ 0&0&\textrm{Var}(\epsilon_{0,i})
 \end{pmatrix}\right).
 $$
When the $i^\textrm{th}$ unit is observed for the treatment status $D=d$, $E(\epsilon_{d,i}) = \mu_{d,i}$, and $\textrm{Var}(\epsilon_{d,i})= \textrm{Variance}(\epsilon_{d,i}) =  \sigma_{d,i}^2$, else $E(\epsilon_{d,i}) = \hat{\mu_{d,i}}$, and $\textrm{Variance}(\epsilon_{i1}) = \hat{\sigma}^2_{d,i}$. Second, the unconditional joint density of $(D^{*},Y(1),Y(0))$ can be computed by marginalizing the latent variable $\Theta$ as
\begin{eqnarray*}
P(D^{*},Y(1),Y(0)|X,Z) &=& \int P(D^{*},Y(1),Y(0)|\Theta,X,Z)p(\Theta)d(\Theta)\\
&\propto&\int \prod_{i} P(D_i^{*}|\Theta_i,X_i,Z_i) \prod_{i\in \{D_i=1\}} P(Y_i(1)|\Theta_i,X_i,\mu_{1,i},\sigma_{1,i}^2) \\
&     &      \   \prod_{i\in \{D_i=0\}} P(Y_i(0)|\Theta_i,X_i,\mu_{0,i},\sigma_{0,i}^2) p(\Theta)d(\Theta)  \\
& = & \int \prod_{i} P(D_i^{*}|\Theta_i,X_i,Z_i) \prod_{i\in \{D_i=1\}} P(Y_i|\Theta_i,X_i,\mu_{1,i},\sigma_{1,i}^2) \\
&     &    \    \prod_{i\in \{D_i=0\}} P(Y_i|\Theta_i,X_i,\mu_{0,i},\sigma_{0,i}^2) p(\Theta)d(\Theta) . 
\end{eqnarray*}
Thus, the integral can be approximated using the Monte Carlo samples from $P(D^{*},Y_{obs}|\Theta,X,Z,
\Omega)$ with observed data via MCMC, when the assumptions of the DPM LIV model are met. The posterior estimate of an average treatment effect can then be computed as
\begin{eqnarray*}
\widehat{\textrm{ATE}}(x) &=& E(E(\Delta |X,\theta, \Omega)) \\
&\approx& E(\frac{1}{M}\sum_{m=1}^M[ (X\beta^{m}_1 + \alpha_1^{m}\theta^{m}+ E(\epsilon_1)) -(X\beta^{m}_0 + \alpha^{m}_0\theta^{m}+E(\epsilon_0)])\\
&=& \frac{1}{n}\sum_{i=1}^n(\frac{1}{M}\sum_{m=1}^M[ (X_i\beta^{m}_1 +
 \alpha_1^{m}\theta_i^{m} + \bar{\mu^m_1}) -(X_i\beta^{m}_0 + \alpha^{m}_0\theta_i^{m}+\bar{\mu^m_0})]), 
\end{eqnarray*}
for MCMC samples $m = 1, \ldots, M$, number of observed data points $n$ and set of model parameters $\Omega$, with $\bar{\mu^m_1} = \frac{1}{k}\sum_{k \in \{D_i = 1\}} \mu_{k1}^m$ and $\bar{\mu^m_0}= \frac{1}{k}\sum_{k \in \{D_i = 0\}} \mu_{k0}^m$ . Similarly, the posterior estimate of an average treatment effect on the treated is given by
\begin{eqnarray*}
\widehat{\textrm{ATT}}(z,x) &=& \int\int E(\Delta|x,\theta,\Omega) \frac{Pr(D=1|\Theta,x,z,\Omega) p(\Theta|\Omega)}{Pr(D=1|x,z,\Omega)} p(\Omega) d\Theta d\Omega \\
&\approx& \frac{1}{M} \sum_{m=1}^M E(\Delta|x,\Theta^{m},\Omega^m) \frac{Pr(D=1|\Theta^m,x,z,\Omega^m)}{Pr(D=1|x,z,\Omega^m)} \\
&=& \frac{1}{M} \sum_{m=1}^M E(\Delta|x,\Theta^{m},\Omega^m) \frac{Pr(D=1|\Theta^m,x,z,\Omega^m)}{\frac{1}{L}\sum_{l=1}^LPr(D=1|x,z,\Theta^{l},\Omega^m)}. 
\end{eqnarray*}
For the proposed model, 
\begin{eqnarray*}
Pr(D_i=1|X_i,Z_i,\Theta_i,\Omega) &=& Pr(D_i^{*} > 0) 
    = Pr(Z_i\gamma+\alpha_D\Theta_i+X_i\beta_D+\epsilon_{Di} > 0) \\
    &=& Pr(\epsilon_{Di} < (Z_i\gamma+\alpha_D\Theta_i+X_i\beta_D)) 
    = \Phi(Z_i\gamma+\alpha_D\Theta_i+X_i\beta_D), 
    \end{eqnarray*}
and $Pr(D=1|X,Z,\Theta,\Omega) = \frac{1}{n} \sum_{i=1}^n\Phi(Z_i\gamma+\alpha_D\Theta_i+X_i\beta_D)$, by the linearity of expectation. Finally, the probability that the difference in outcome is greater than threshold $H$ can be estimated as
$\widehat{\textrm{PB}}(x) = 
\frac{1}{M}\sum_{m=1}^m Pr(Y(1)-Y(0) > H |X,\Theta^{(m)},\Omega^{(m)})$, and by noting that $Y(1) \perp Y(0)| \Theta$.

\section{Simulation Study}
\label{sec::simulation}
We explore the performance of our proposed model, comparing it to an existing Bayesian model, the latent factor model with Normal noise (Normal LIV) \citep{heckman2014treatment}, and a non-Bayesian model, the two-stage least squares (2SLS) \citep{angrist1996identification} method, via simulations. Continuous potential outcomes with binary treatment and a binary instrument similar to our application are considered. The data generation process follows Equation \ref{eq::heckmanIV_Latent}.  The matrix of covariates, $X$, is simulated with three columns such that $X_1$ and $X_2$ are continuous and $X_3$ is binary. Additionally, $\beta_0^0$ is fixed at 90, $\beta_0$ is fixed at  $\{-0.5,1.5,0\}$, $\beta_1^0$ is fixed at 100, $\beta_1$ is fixed at $\{-0.5,1.5,10\}$, $\beta_D^0$ is fixed at -0.5, and $\beta_D$ is fixed at $\{0,0,1\}$. $\Theta$ is generated from Normal$(0, s)$ distribution, with $s = \{0.1,0.5,1,\sqrt{10}, 10\}$ as different considerations for standard deviations. These parameters are chosen so that the simulated data imitate our data application with respect to the skewed distribution of the outcome model along with a heterogeneous treatment effect based on covariate $X_3$. Approximately 30\% of the patients are {\em treated} in the simulation setting to reflect the proportion undergoing radial artery access in our PCI cohort.

The distribution of the errors in potential outcomes, $\epsilon_0$ and $\epsilon_1$, is varied to quantify the effect of different error models on the estimates of the ATE and CATE. The error of the treatment model, $\epsilon_D$, is simulated from Normal(0, 1). The data are simulated under both weak and strong instruments as measured by the coefficient $\gamma$, and different correlation coefficients as specified by coefficients $\alpha_0$, $\alpha_1$ and $\alpha_D$.  Below, we present results from the simulations where the errors of the potential outcomes $\epsilon_0$ and $\epsilon_1$ are generated from Gamma(shape=3, rate=1/10), distributions with strong instrument ($\gamma = 1.5$), $\alpha_0 = 0.1$, $\alpha_1 = 0.1$ and $\alpha_D = 0.2$, and $\Theta \sim \text{Normal}(0,0.1)$. Finally, $Y_{obs} = D Y + (1-D) Y$ is used for fitting the model and estimating treatment effects. Results for the remaining conditions are shown in the Supplemental Material.

Data are simulated under varying sample sizes $n$. For each $n$, we generate and fit the data for 50 replications. Relatively non informative prior of Normal(0, 100) is used on the coefficients in outcome and treatment models. The latent factor $\Theta$ is assumed to have Normal(0, 1) prior distribution for identification and the hyperpriors on the stick breaking formulation of the DPM priors are specified as $c \sim\textrm{Gamma}(a, b)$ with $a=1, b=1$, $\tau \sim \textrm{InverseWishart}(\nu, \Psi)$ with $\nu=1$ and $\Psi^{-1} = 5$, $\omega \sim  \textrm{Normal}(m, K)$, $m \sim \textrm{Normal}(0, 1)$ and $K \sim \textrm{InverseGamma}(1,10)$. For each fit, 20,000 MCMC samples were drawn with burn in of 5,000 and every 10$^{th}$ draw retained for inference. Mean bias, mean width of the posterior 95\% credible interval, and percentage coverage of 95\% posterior credible intervals were computed for ATE and CATE, as defined in Section \ref{sec::effects}. The simulation results are reported in Table \ref{tab::ATE}. 
\begin{table}[!ht]
\centering
\begin{tabular}{cccccc}
\hline
 {\bf Causal}            &        &             & {\bf Absolute} & \multicolumn{2}{c}{\bf 95\% Credible Interval}  \\ \cline{5-6}
{\bf Parameter} & \bf n  & \bf Method &  {\bf Bias} & \bf Width & \bf \% Coverage \\
\hline
{\textbf{ATE}} & 100 & DPMLIV  & 2.9	& 18.4	&   98	\\
(true = 14) 	&	&  NormalLIV &3.1		& 26.6 	&  100	\\
		& 	&	2SLS  & 5.4	   	&  30.1	&	98 \\
\cline{2-6}
		&500    & DPMLIV		& 1.4	&  9.2	& 100	\\
 		& 	& NormalLIV		&	1.6       &   13.5	& 100   \\
		& 	& 2SLS  			&  2.9       & 16.9 	  & 	96	\\
\cline{2-6}
	&	2000   	&  DPMLIV  	& 0.8	& 4.5  	& 98	\\
 		&	& NormalLIV 		& 0.9	&  7.4	&    100\\
		& 	& 2SLS  		&1.9 	& 8.7	&	96\\
\hline
{\textbf{CATE}}($X_3=1$) &100		& DPMLIV	& 3.0 	& 21.3	&100	\\
(true = 18) 		& & NormalLIV			& 4.2	& 36.3	& 100	\\
		& &2SLS  			&       5.8& 	33.8	& 96\\
\cline{2-6}
	&500   		& DPMLIV    & 1.8       &   10.7	& 98	\\
		&	& NormalLIV 		&3.5&	        17.1  &    	96 \\
		&	&2SLS  	 	&    5.6     &26.8  &  92\\
\cline{2-6}
	&2000   & DPMLIV  		&	 0.8 &  5.4& 100 \\
	&	& NormalLIV 		& 	1.8&  	9.6&  100 \\
	&	& 2SLS  		& 	 2.7& 12.1    &	98 \\
\hline
\end{tabular}
\caption{{\bf Mean absolute bias, mean width of posterior 95\% credible interval and coverage of 95\% credible interval of estimated average treatment effect (ATE) and conditional average treatment effect (CATE($X_3=1$)) under Gamma(3, 0.1) errors.} Results for varying sample sizes for three different methods. True value of $\beta_0^0$ is fixed at 90, $\beta_0$ is fixed at  $\{-0.5,1.5,0\}$, $\beta_1^0$ is fixed at 100, $\beta_1$ is fixed at $\{-0.5,1.5,10\}$, $\beta_D^0$ is fixed at -0.5, $\beta_D$ is fixed at $\{0,0,1\}$, and $\Theta$ is generated from Normal(0, 0.1) distribution. The errors $\epsilon_0$ and $\epsilon_1$ are generated from Gamma(3, 0.1) distributions; $\gamma = 1.5$; $\alpha_0 = 0.1$, $\alpha_1 = 0.1$ and $\alpha_D = 0.2$.} 
\label{tab::ATE}
\end{table}

Sensitivity analysis for the hyper priors on the concentration parameter $c$ and variance $\tau$ on DPM priors is further conducted to investigate whether and to what degree the estimates from the proposed model are sensitive to the prior assumptions. We use simulated data with $n = 2000$, strong IV, and various specifications of the hyper parameters. Results, as shown in the Supplemental Material, suggest that bias and 95\% credible interval width of the estimated treatment effects are agnostic to different hyper prior assumptions. However, we observe some variation on the percentage coverage of 95\% posterior credible intervals for the hyper parameter for $\tau$.

From these simulations, we observe that incorporating DPM priors on the errors of the potential outcome models decreases overall bias and width of the the credible interval when estimating the ATE and CATE, particularly when the errors of the potential outcome deviate from the Normal distribution. Thus, among the different models considered, the DPM LIV performs the best, even when the underlying errors have Normal symmetric distribution (as demonstrated by the simulations in the Supplemental Material). The performance of the widely used 2SLS method improves with increases in sample size and the strength of the instrument while computing the ATE. But the practical use of the 2SLS remains limited because we do not have a way to compute the CATE without stratifying the samples and fitting different models on each sample.

\section{Data Analysis} 
\label{sec::dataintro}

\subsection{Availability of a Valid Instrument and Sensitivity Analysis}
\label{ssec::instrument}
\begin{figure}[!ht]
\centering
\includegraphics[height=3in,width=3in]{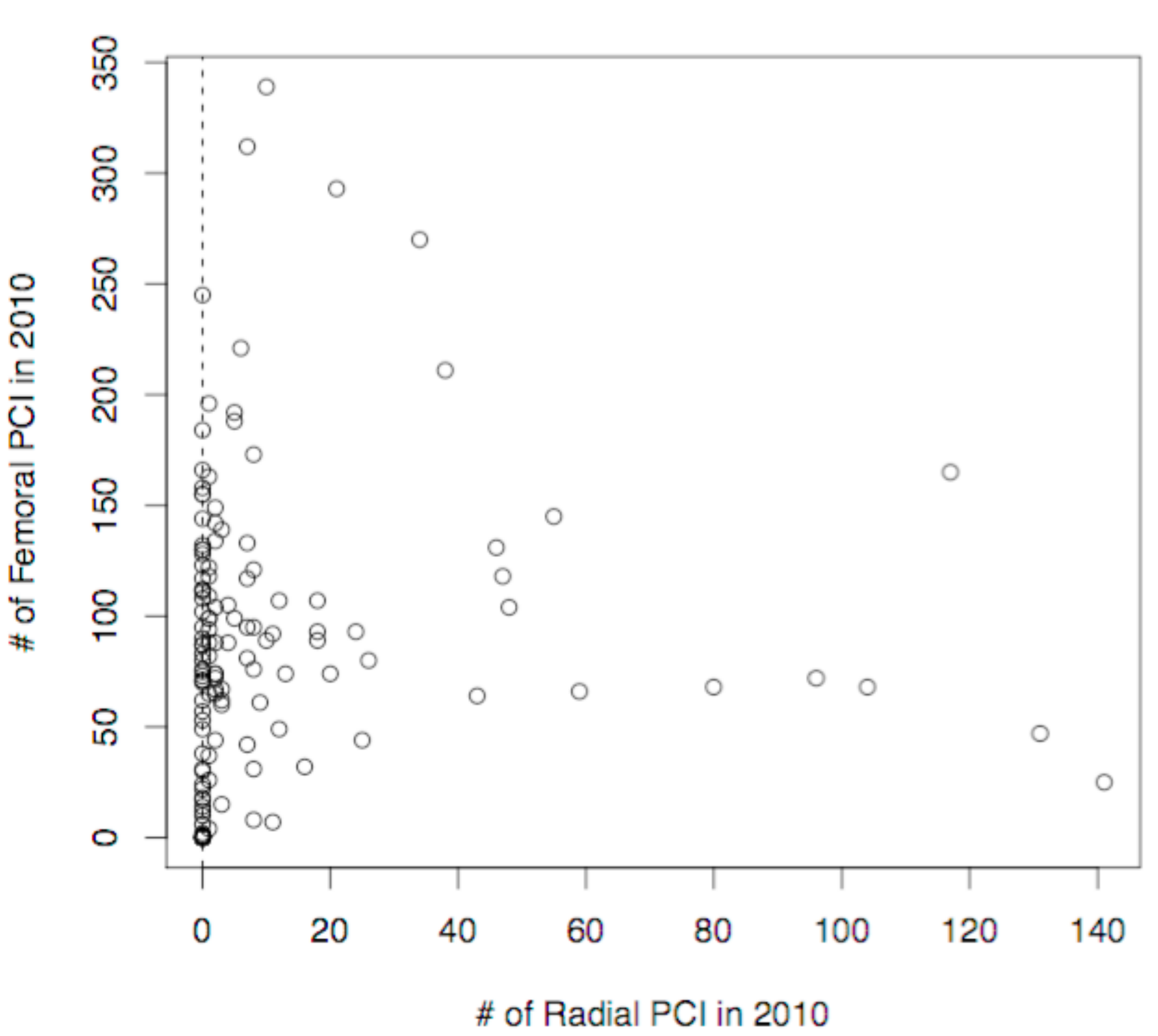}
\caption{{\bf Scatter plot of femoral arterial access PCI and radial arterial access PCI performed by each physician in 2010.} Dotted vertical line represents surgeons who had not performed any radial surgery in the previous year. Any PCI that was performed by the physicians to the right of the vertical line in the year 2010 will have an IV = 1; otherwise the physician will have IV = 0.}
\label{fig::RadialvsFemoral}
\end{figure}

Our choice of an IV is motivated by prior experience of physicians performing radial arterial access PCI. Radial arterial access PCI is a relatively new procedure with a steep learning curve for physicians \citep{ball2011characterization}. Physicians who have traditionally performed femoral arterial arterial access PCI are less likely to perform a radial approach \citep{ball2011characterization}. Figure \ref{fig::RadialvsFemoral} shows a scatter plot of number of radial and femoral arterial access PCIs performed by the physicians during 2010. The majority of the physicians have performed femoral arterial access PCIs only. Using this information, a binary instrument that is 1 if a physician has any prior experience on performing radial arterial access PCI in the year 2010 and zero otherwise, is created for use in our 2011 cohort. We observe that the patients for whom the IV is 1 are 24\% more likely to receive radial arterial access PCI than those with IV=0. 

Validity of the proposed instrument is evaluated with sensitivity analysis by implementing the tests suggested in \citet{SIM:SIM6128}. We first assess the strength of the IV using an F-statistic that is computed by adding an IV to the reduced first-stage model for the treatment, after including the measured confounders $X$ \citep{bound1995problems, SIM:SIM6128}. As a general rule F-statistic greater than 10 suggests a strong IV. By fitting the full and reduced logistic regression in the treatment equation with and without the instrument, an F-statistic of 665 is obtained, suggesting a strong instrument. We then estimate proportion of compliers \citep{ertefaie2017tutorial, SIM:SIM6128} by calculating the difference of the treatment assignment rate among subjects across the values of the IV i.e., $P(D=1|Z=1)-P(D=1|Z=0)$. The estimated proportion of compliers in our analysis is $P(D=1|Z=1)-P(D=1|Z=0) = 0.298-0.055 = 0.243$, indicating a reasonable effective sample size \citep{ertefaie2017tutorial}.

Next, the exclusion restriction assumption made in Assumption \ref{ass:er} is assessed. Recall that the 1-year TVR rate should be unaffected by the treatment choice of radial or femoral arterial access PCI. As \citet{SIM:SIM6128} suggest, if an IV is not associated with a treatment-unaffected outcome, we are more confident that the exclusion restriction assumption holds. The differences in TVR at different levels of IV, for overall sample and when stratified by sex is less than 1\%, as shown in Table \ref{tab::IVprop}, suggesting that the assumption may hold in our analysis. 
\begin{table}[!ht]
 \centering
 \begin{tabular}{r rr rr rr}
   \hline
  & \multicolumn{2}{c}{\bf Males}  &  \multicolumn{2}{c}{\bf Females} & \multicolumn{2}{c}{\bf Overall } \\ \cline{2-7}
{\bf Radial PCI in 2010} & \bf None & \bf $>$ 1  & \bf None & \bf $>$ 1  & \bf None & \bf $>$ 1  \\ \hline
No. PCI Admissions & 1,563 &3,903  &   716  &1,781 & 2,279 &  5,684 \\  \hline
  \hline
\bf Radial PCI in 2011 \% & 5.8	&	32.1 &4.5	& 27.5	 & 5.5 & 29.8 \\ 
\hline
{\bf In-Hospital Outcomes} &    & 	 &	&  &	&  \\
 Mean Total Charges, \$ &   52,790	& 52,205	& 50,0311	& 50,915    & 52,011 & 51,798 \\ 
 Vascular Complications, \% & 2.8	& 2.5	&	6.6 & 6.3	& 4.8 & 4.5 \\ 
 \hline
 30-Day Major Bleed, \% &	2.7&	2.5& 3.8	& 5.5	& 3.9 & 4.4 \\ 
 \hline
{\bf 1-Year Outcomes, \%} &  &  &	&	&	& \\
Major Bleed & 7.9	&	7.6&	12.2& 12.9 	& 12.5 & 12.1 \\ 
 \textit{TVR} &	\textit{6.8} & \textit{6.9}	& \textit{6.9}	& \textit{7.4}	& \textit{7.2} & \textit{7.9} \\ 
\hline
\end{tabular}
\caption{{\bf Treatment and unadjusted mean outcomes stratified by IV \& sex.} Bleed defined by an in-hospital admission for major bleeding.}
\label{tab::IVprop}
 \end{table}

The independence assumption of the IV and unmeasured confounders cannot be confirmed using observed data. However, examination of the balance of measured covariates between values of the IV may provide insight about the validity of this assumption \citep{ertefaie2017tutorial, SIM:SIM6128}. We report the balance of measured covariates between the levels of the IV in the Supplemental Material and observe that the confounders are fairly balanced in our application.

\subsection{Model Estimation and Results}
\label{sec::analysis} 
The DPM LIV model from Equations \ref{eq::heckmanIV_Latent} and \ref{eq::IV_DPM} is fitted on the observed data to compare the effectiveness of radial arterial access compared to femoral arterial access PCI in reducing hospitalization charges at discharge.
Relatively noninformative prior of Normal(0, 100) is used on the coefficients in outcome and treatment models. The latent factor $\Theta$ is assumed to have Normal(0, 1) prior distribution for identification and the hyperpriors on the stick breaking formulation of the DPM priors are specified as $c \sim\textrm{Gamma}(a,b)$ and $\tau \sim \textrm{InverseWishart}(\nu, \Psi)$. We conducted sensitivity analysis using both informative and noninformative prior specifications on $c$ and $\tau$. The results from the sensitivity analysis are reported in the Supplemental Material. The estimated posterior median of each treatment effect with different prior specifications are within \$100 of each other, with very small difference on the posterior 95\% credible intervals. The sensitivity analysis suggests that our findings are robust to the hyper-parameter specifications on the DPM priors.

To illustrate our findings, we continue with the assumption that $a=1, b=10$, $\nu=2$ and $\psi^{-1} = 50$. Ten thousand MCMC samples were drawn, and after 2,000 burn in and 10 thinning we retain an effective sample size of 800. Convergence of the chains was confirmed by assessing trace plots and ensuring that the Gelman-Rubin statistics \citep{gelman2014bayesian} were below 1.1. The estimated posterior median along with 95\% credible interval around the median of the treatment effects are reported in Table \ref{tab::result}. The threshold $H$ for estimating the PB$(x)$ is specified as 10\% of the median overall observed charges (\$41,387).  Plots of the posterior distribution of the treatment effects are shown in the Supplemental Material.
\begin{table}[!ht]
\centering
\begin{tabular}{c c c}
\hline
\bf Estimated Treatment Effect & \bf Posterior Median & \bf 95\% Credible Interval \\
\hline
\bf ATE &	-\$3,849 &	(-\$5,099,	-\$2,632)	\\
\bf CATE, Female &   -\$2,815 & 	(-\$4,211,	-\$1,458)		\\
\bf CATE, Male & -\$4,321 &	(-\$5,664,	-\$3,003)	\\
\bf ATT, Radial & 	-\$3,841 &	(-\$5,158,	-\$2,665)		\\
\bf PB & $ 0.41$  & ($0.38, 0.44$) \\
\hline
\end{tabular}
\caption{{\bf Posterior estimates of causal effects in the PCI data.} Estimates of the average treatment effect (ATE), sex specific average treatment effects (CATE, Female and CATE, Male), and the average treatment effect on the treated (ATT) of the difference in hospitalization charges of radial and femoral arterial access PCI reported. The estimated probability that the difference in charges between femoral and radial arterial access PCI is greater than $H$, i.e. Pr(Charges Femoral $-$ Radial $> H$ ), or simply PB also shown. Here, $H$= 10\% of median observed charges = 4,139.}
\label{tab::result}
\end{table}

Radial arterial access PCI reduces the overall hospitalization charges compared to femoral arterial access PCI with a mean reduction of around \$3,850. Furthermore, the reduction is about \$1,500 more in male patients compared to the charges for female patients. Estimated ATE between radial and femoral arterial access PCI is -\$2,815 in female patients compared to -\$4,321 in male patients. The estimated ATT for radial arterial access PCI is similar to the estimated ATE.  The posterior 95\% credible interval for the ATT is wider and overlaps with that of the overall ATE. Finally, the estimated probability that the difference in charges for femoral and radial arterial access PCI exceeds 10\% of the median observed charges is 41\%, further supporting our finding that radial arterial access is more effective in reducing hospitalization charges compared to femoral arterial access.

Our results suggest evidence of heterogeneity in the effects of radial arterial access compared to femoral arterial access PCI on hospitalization charges, based on sex of the patients. However, the heterogeneity in the treatment group compared to the overall sample is not very pronounced. Compared to the unadjusted difference in charges, the estimated difference in average charges for women is considerably higher after accounting for confounders and adjusting for unobserved confounders using IV analysis. The overall ATE after adjusting for the instrument is higher than that estimated by only adjusting for observed confounders.

\section{Discussion}
\label{sec::conclusion}
In this paper, we studied heterogeneous treatment effects of different arterial access site strategies for PCI in reducing procedure related hospitalization charges at discharge, and proposed a novel Bayesian method to extend existing methodology for estimating treatment effects, when there is selection bias. Our new method relaxes assumptions of Normal errors in outcome models and homogeneous selection in treatment models that are often made in latent index variable models. Simulations established that our method performs better than existing methods and has reasonable frequentist properties. 

Our overall finding is that radial arterial access PCI reduces hospitalization charges compared to femoral arterial access PCI. This finding is to some extent consistent with that of RCTs conducted previously, which have found that radial arterial access PCI is better than femoral arterial access PCI in reducing major bleeding and in-hospital complications. However, our additional finding about the heterogeneous effects in male patients and female patients for hospitalization charges at discharge after PCI is novel. We found a reduction in charges after radial arterial access compared to femoral arterial access PCI to be higher for male patients than for female patients. This finding of differences in charges based on sex and different arterial access PCIs is in some way contrary to the findings in RCTs, which have shown that the radial arterial access PCI is more effective in female patients compared to male patients. We have an interesting result which certainly requires further investigation from practitioners and cardiac health researchers. 

Methodologically, there are many directions to expand on the proposed work. While the main focus of this paper is on continuous outcomes, our work can be extended to other types of outcomes using generalized linear models. Our results assume linear relationship between the outcome and the covariates. There are many situations when this assumption of linearity might not hold. It will be important to explore methods that extend such restriction in the latent index modeling framework. We have also largely ignored the hierarchical structure of the data where the procedures are nested within hospitals. It is essential to expand our work to account for such hierarchical structures. Further, we rely on many assumptions to make causal inference. Some of these assumptions are either testable or can be verified by sensitivity analysis as shown in this paper. However, there are assumptions that cannot be empirically tested or verified. It will be important to expand this work to include sensitivity analysis for such assumptions.

\section*{Acknowledgement}
We are indebted to the Massachusetts Department of Public Health (MDPH) and the Massachusetts Center for Health Information and Analysis Case Mix Databases (CHIA) for the use of their data. Funded by R01-GM111339 and U01-FD004493.

\bibliographystyle{asa}
\bibliography{BCER_Citation}

\appendix
\section{Appendix}

\listoffigures
 
\listoftables

\begin{table}[ht]
\centering
\begin{tabular}{llrr}
  \hline
 &\bf Covariates &\bf Radial &\bf Femoral \\ 
&		& (n = 1,823) & (n=6,140) \\
  \hline
\hline
\bf Demographics,\% &  &	&	\\
	&  $^\dag$Mean Age, yrs & 63.91 & 65.96 \\ 
  	&  Caucasian & 92 & 92 \\ 
 	&  Male & 74 & 70 \\ 
  	&  Government Health Insurance & 46 & 51 \\ 
\hline
\hline
\bf Comorbidities, \% &	&	&	\\
  	& Past or Current Smoker & 25 & 24 \\ 
	& Hypertension & 77 & 79 \\ 
	& Current Dialysis & 1 & 2 \\ 
  	& Diabetes & 32 & 32 \\ 
  	& Chronic Lung Disease & 13 & 14 \\ 
  	& Dyslipidemia & 82 & 83 \\ 
  	& Family History  &   &   \\
	& Coronary Artery Disease & 32 & 29 \\ 
  	& Prior Myocardial Infarction & 24 & 28 \\ 
  	& Prior Heart Failure & 8 & 12 \\ 
  	& Prior PCI & 28 & 29 \\ 
  	& Prior CABG & 8 & 15 \\ 
  	& Prior Cerebrovascular Disease & 9 & 10 \\ 
  	& Prior Peripheral Artery Disease & 11 & 12 \\ 
  	& Peripheral Vascular Disease & 5 & 5 \\ 
  	& Congestive Heart Failure & 10 & 14 \\ 
\hline
\hline
\bf Cardiac Presentation, \% &	&	&	\\
  	& Cardiogenic Shock & 1 & 3 \\ 
 	& Elective Status & 31 & 28 \\ 
  	& Urgent Status & 52 & 46 \\ 
  	& Emergent or Salvage Status & 16 & 26 \\ 
  	& Left Main Disease & 4 & 7 \\ 
  	& LAD Artery Stenosis $>70$\% & 57 & 61 \\ 
  	& Unstable Angina & 35 & 29 \\ 
  	& Stable Angina & 16 & 17 \\ 
  	& Non-STEMI & 26 & 23 \\ 
  	& STEMI & 15 & 24 \\ 
  \hline
\end{tabular}
\caption{Baseline covariates that were adjusted for in the analysis for radial and femoral PCI. Note that the patients receiving radial arterial access PCI have less severe conditions compared to those receiving femoral arterial access PCI, hence adding to the evidence that healthier patients usually get radial arterial access PCI. $\dag$ mean age reported; CABG = coronary artery bypass grafting; LAD = left anterior descending; STEMI = ST-Elevated Myocardial infarction.}

\label{table::Confounders}
\end{table}

\begin{table}[!ht]
\centering
\begin{tabular}{llcc}
  \hline
&\bf  Covariates 	&	\bf (IV = 0) & \bf ( IV = 1) \\ 
&	& (n=2,279) & (n=5,684) \\
  \hline
\hline
\bf Demographics, \% &		&		& \\
	& $^\dag$Mean Age, yrs & 65.31 & 65.56 \\ 
  	& Caucasian & 94 & 91 \\ 
  	& Male & 71 & 70 \\ 
  	& Government Health Insurance  & 51 & 50 \\ 
\hline
\hline 
\bf Comorbidities, \% & 	&	& \\  
  	& Past or Current Smoker & 25 & 24 \\
	& Hypertension & 80 & 78 \\ 
  	& Dyslipidemia  & 84 & 82 \\ 
 	& Family History &	& \\
	& Coronary Artery Disease & 32 & 29 \\ 
 	& Prior Myocardial Infarction & 26 & 28 \\ 
  	& Prior Heart Failure & 10 & 11 \\ 
  	& Prior PCI & 29 & 29 \\ 
  	& Prior CABG & 13 & 13 \\ 
  	& Prior Cerebrovascular Disease & 10 & 10 \\ 
  	& Prior Peripheral Artery Disease & 12 & 12 \\ 
 	& Current Dialysis & 2 & 2 \\ 
  	& Diabetes & 30 & 33 \\  
  	& Chronic Lung Disease & 13 & 14 \\ 
    & Peripheral Vascular Disease & 4 & 5 \\ 
    & Congestive Heart Failure & 12 & 14 \\ 
 \hline
\hline
\bf Cardiac Presentation, \%&	&	& \\
  &	Cardiogenic Shock & 3 & 2 \\ 
  &	 Elective Status & 30 & 28 \\ 
  &	 Urgent Status & 45 & 49 \\ 
  &	 Emergent or Salvage Status & 26 & 23 \\ 
  &	 Left Main Disease & 6 & 6 \\ 
  &	 LAD Artery Stenosis $>70$\% & 60 & 60 \\ 
  &	 Unstable Angina & 28 & 32 \\ 
  &	 Stable Angina & 18 & 16 \\ 
  &	 Non-STEMI & 22 & 24 \\ 
  &	 STEMI & 24 & 21 \\ 
   \hline
\end{tabular}
\caption{Baseline covariates at different levels of instrument to investigate the independence assumption of the IV and unmeasured confounders. We observe that the confounders are fairly balanced, suggesting that the assumption of the independence of the IV and unmeasured confounders is valid.$\dag$ mean age reported; CABG = coronary artery bypass grafting; LAD = left anterior descending; STEMI = ST-Elevated Myocardial infarction.}
\label{table::instrument}
\end{table}

\begin{figure}
\includegraphics{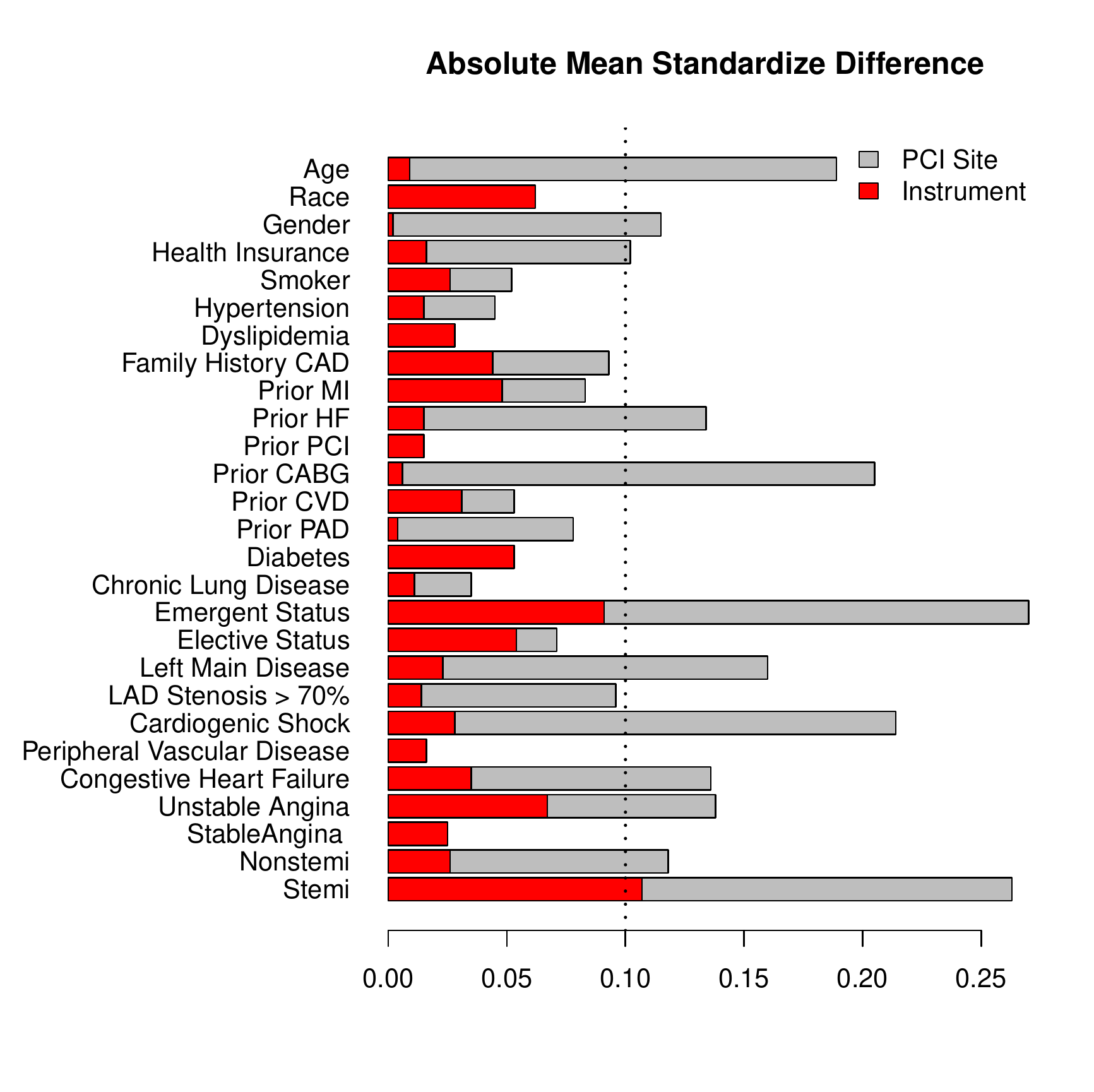}
\caption{Overlaid barplots of absolute mean standardized differences between, i.) patients with different arterial access PCIs (in grey), ii.) patients at different levels of instruments (in red), for each observed confounder. Note that the mean standardized difference is usually larger when comparing confounders between radial and femoral arterial access PCI compared to that between different levels of the instrument, implying that the covariates are more balanced at different levels of instruments. CAD = coronary artery disease; MI = myocardial infarction; HF = heart failure; CABG = coronary artery bypass grafting; CVD = cerebrovascular disease; PAD = peripheral artery diseaseLAD = left anterior descending; STEMI = ST-Elevated Myocardial infarction.}
\label{}
\end{figure}

\begin{table}[!ht]
\centering
\begin{tabular}{lccc}
\hline
\bf Hyperparameters & \bf Treatment Effect & \bf Posterior Median & \bf 95\% Credible Interval \\
\hline

$\bf (a, b) =  (10, 1)$ \& &\bf ATE &	-\$3,849 &	(-\$5,099,	-\$2,632)	\\
$\bf (\Psi^{-1}, \nu) = (50,2)$ &\bf CATE, Female &   -\$2,815 & 	(-\$4,211,	-\$1,458)		\\
					&\bf CATE, Male & -\$4,321 &	(-\$5,664,	-\$3,003)	\\
				&\bf ATT, Radial & 	-\$3,841 &	(-\$5,158,	-\$2,665)		\\
				&\bf PB & $ 0.41$  & ($0.38, 0.44$) \\

\hline


$\bf (a, b) = (10, 1)$ \& &\bf ATE & -\$3,895 &	(-\$5,081,	-\$2,758)	\\
$\bf (\Psi^{-1}, \nu) = (1,4)$ &\bf CATE, Female &   -\$2,949	& (-\$4,268,  -\$1,598)	\\
						&\bf CATE, Male & 	-\$4,310 &	(-\$5,574,	-\$3,136)	\\
						&\bf ATT, Radial & -\$3,887 & (-\$5,103,	-\$2,678)	\\
						&\bf PB & $0.41$  & ($0.37, 0.44$) \\
\hline
$\bf (a, b) = (1, 1)$ \& & \bf ATE & -\$3,910 &	(-\$5,309,	-\$2,629) \\
$\bf (\Psi^{-1}, \nu) = (50,2)$ & \bf CATE, Female & -\$2,830 &	(-\$4,312,	-\$1,394) \\
				&\bf CATE, Male &-\$4,379 &	(-\$5,797, -\$3,116) \\
				&\bf ATT, Radial & -\$3,890 &	(-\$5,309,	-\$2,575) \\
				&\bf PB & $0.41$  & ($0.38, 0.44$) \\
\hline

$\bf (a, b) = (1, 1)$ \&  &\bf ATE &-\$3,856 &	(-\$5,136,	-\$2,696) \\
$\bf (\Psi^{-1}, \nu) = (1,4)$&\bf CATE, Female & -\$2,947 &	(-\$4,436,	-\$1,635) \\
				&\bf CATE, Male & -\$4,321 &	(-\$5,544,	-\$3,105) \\
				&\bf ATT, Radial & -\$3,842 &	(-\$5,184,	-\$2,711) \\
				&\bf PB & $ 0.41$  & ($0.38, 0.44$) \\
				\hline
\end{tabular}
\caption{{\bf Sensitivity Analysis: Posterior estimates of causal effects in the PCI data under varying hyperprior specifications on $c \sim \text{Gamma}(a, b)$ and $\tau \sim \text{InverseWishart}(\Psi^{-1}, \nu)$.} Estimates of the average treatment effect (ATE), sex specific average treatment effects (CATE, Female and CATE, Male), and the average treatment effect on the treated (ATT) of the difference in hospitalization charges of radial and femoral arterial access PCI reported. The estimated probability that the difference in charges between femoral and radial arterial access PCI is greater than $H$, i.e. Pr(Charges Femoral $-$ Radial $> H$ ), or simply PB also shown. Here, $H$= 10\% of median observed charges = 4,139. Hyperparameters $(a, b)$ and $(\Psi^{-1}, \nu)$ are presented in the order of being noninformative to informative. }
\label{tab::result}
\end{table}


\begin{table}[ht]
\centering
\begin{tabular}{ccccc}
  \hline
   {\bf Causal} &  & {\bf Absolute} & \multicolumn{2}{c}{\bf 95\% Credible Interval} \\ \cline{4-5}
{\bf Parameter} &  \bf Method &  {\bf Bias} & \bf Width & \bf \% Coverage \\
\hline
{\textbf{ATE}}& DPMLIV & 0.07 & 0.45 & 100 \\ 
 (true = 15) 		& NormalLIV & 0.07 & 0.49 & 100 \\ 
  		& 2SLS & 0.29 & 2.34 & 100 \\ 
   \hline
{\textbf{CATE}}($X_3=1$)  	& DPMLIV & 0.07 & 0.51 & 98 \\ 
 (true = 20)	& NormalLIV & 0.03 & 0.62 & 100 \\ 
 	& 2SLS & 0.12 & 0.68 & 100 \\ 
\hline
\end{tabular}
\caption{{\bf Simulation results under Normal outcome errors.} Mean absolute bias, mean width of posterior 95\% credible interval and coverage of 95\% credible interval of estimated ATE and CATE for $n = 2000$. True value of $\beta_0^0$ is fixed at 90, $\beta_0$ is fixed at  $\{-0.5,1.5,0\}$, $\beta_1^0$ is fixed at 100, $\beta_1$ is fixed at $\{-0.5,1.5,10\}$, $\beta_D$ is fixed at $\{0,0,1\}$ and $\Theta$ is generated from Normal(0,0.1) distribution; errors on the potential outcomes $\epsilon_1$ and $\epsilon_0$ are generated from Normal (0,0.5) distributions with strong instrument ($\gamma$=1.5), and $\alpha_0 = 0.1$, $\alpha_1 = 0.1$ and $\alpha_D = 0.2$.}

\end{table}

%

\begin{table}[!ht]
\centering
\begin{tabular}{cccccc}
\hline
 &  & & {\bf Absolute} & \multicolumn{2}{c}{\bf 95\% Credible Interval}  \\ \cline{5-6}
\bf n  & $\gamma$& \bf Method &  {\bf Bias} & \bf Width & \bf \% Coverage \\
\hline
100&0.5 & DPMLIV  &	0.5	&10	&100	\\
&	&  NormalLIV &	1.2	& 9	&100	\\
	& &2SLS  & 	   	2.0&  31	&	100\\
\hline
500   & & DPMLIV		&0.6	&  2.0	& 100	\\
 	& & NormalLIV		&	    0.7    &  2.7 	&  100  \\
	& & 2SLS  			&      3.0   & 	12.5  & 	100	\\
\hline
2000   &	&  DPMLIV  			&	0.1&  1.1	& 100	\\
 	&	& NormalLIV 		& 0.1	&  1.4	&     100\\
	& 	& 2SLS  			& 0.9	& 5	&	100\\
\hline \hline
100		&1.5 & DPMLIV	& 0.5	&5.0	& 100	\\
	& & NormalLIV		& 1.0	& 	6.3	& 100	\\
	& &2SLS  			&       1.7& 	9.8	& 100\\
\hline
500   	&	& DPMLIV  			&      0.7 &    1.9 	&98	\\
	&	& NormalLIV 		&	          0.8&  2.4  	&   96  \\
	&	&2SLS  	 	&         2.3 &  4.2&  18	\\
\hline
2000   &	& DPMLIV  				&	0.1  &0.6	& 	96\\
	&	& NormalLIV 				& 	0.1&  0.9	&  100 \\
	&	& 2SLS  				& 	0.2&   1.6 &	100 \\
\hline
\end{tabular}
\caption{{\bf Simulation result under mixture normal outcome errors: ATE.} Mean absolute bias, mean width of posterior 95\% credible interval, and coverage of 95\% credible interval of estimated ATE under weak and strong instruments and varying sample sizes for three different methods. True value of $\beta_0^0$ is fixed at 90, $\beta_0$ is fixed at  $\{-0.5,1.5,0\}$, $\beta_1^0$ is fixed at 100, $\beta_1$ is fixed at $\{-0.5,1.5,10\}$, $\beta_D$ is fixed at $\{0,0,1\}$; $\Theta$ is generated from Normal(0,0.1) distribution; $\alpha_1=0.01$, $\alpha_0=0.01$, $\alpha_D=0.01$; $\epsilon_1$ and $\epsilon_0$ are generated from a 4-component mixture of Normal distributions with weights $w=(0.15,0.4,0.25,0.05)$, mean parameter vector $\mu = (-0.1,1,1,10)$, and variance vector $v = (0.01,0.1,0.1,10)$; true ATE is 14.}
\label{tab::ATE}
\end{table}

\begin{table}[!ht]
\centering
\begin{tabular}{cccccc}
\hline
 &  & &{\bf Absolute} & \multicolumn{2}{c}{\bf 95\% Credible Interval}  \\ \cline{5-6}
\bf n  & $\gamma$& \bf Method &  {\bf Bias} & \bf Width & \bf \% Coverage \\
\hline
100 & 0.5	&DPMLIV  		&	        0.27  & 8.6	          &100	\\
	&	& NormalLIV	& 0.75	& 5.4 	&100	\\
	& 	&	2SLS  		&  4.67	&  38.5	&100\\
\hline
500 &	& DPMLIV	 	&  0.5	& 	2.6&	100\\
  	& &NormalLIV 	& 0.4	& 2.3	&    100 \\
	&  & 2SLS  			&2.2 	& 25.4	&	100\\
\hline
2000&   & DPMLIV  	 	& 0.07	& 0.8	&	100\\
	&	&  NormalLIV  	& 0.15 	&1.6	&    100 \\
	&	&2SLS  				& 2.02	& 	8.86&	100\\
\hline \hline
100 &1.5	&	DPMLIV 	& 0.09	& 1.7	& 100	\\
	& & NormalLIV	& 0.2	& 	2.2& 100	\\
	& &2SLS 	& 1.1	& 5.8	& 100\\
\hline
500   	&		 &DPMLIV  	 	&0.08	&	1.4 &  100	\\
	&	& NormalLIV& 	0.27&  1.97	&     100\\
	&	&2SLS  			& 0.8	& 3.19	&	100\\
\hline

2000   	&	&DPMLIV  		& 0.09	&  0.4	& 100	\\
	&	& NormalLIV  	& 	   0.04&	0.99 &  100     \\
	&	&2SLS  				& 0.67 & 1.75 	& 100	\\
\hline
\end{tabular}
\caption{{\bf Simulation result under mixture normal outcome errors: CATE.} Mean absolute bias, mean width of posterior 95\% credible interval, and coverage of 95\% credible interval of estimated CATE ($\textrm{CATE}|X_3=1$) under weak and strong instruments and varying sample sizes for three different methods. True value of $\beta_0^0$ is fixed at 90, $\beta_0$ is fixed at  $\{-0.5,1.5,0\}$, $\beta_1^0$ is fixed at 100, $\beta_1$ is fixed at $\{-0.5,1.5,10\}$, $\beta_D$ is fixed at $\{0,0,1\}$; $\Theta$ is generated from Normal(0,0.1) distribution; $\alpha_1=0.01$, $\alpha_0=0.01$, $\alpha_D=0.01$; the errors $\epsilon_1$ and $\epsilon_0$ are generated from a 4-component mixture of Normal distributions with weights  $w= (0.15,0.4,0.25,0.05)$, mean parameter vector $\mu = (-0.1,1,1,10)$, and variance vector $v = (0.01,0.1,0.1,10)$; true CATE is 19.}
\label{tab::CATE}
\end{table}

\begin{table}[!ht]
\centering
\begin{tabular}{cccccc}
\hline
 &  & &{\bf Absolute} & \multicolumn{2}{c}{\bf 95\% Credible Interval}  \\ \cline{5-6}
\bf n  & $\gamma$& \bf Method &  {\bf Bias} & \bf Width & \bf \% Coverage \\
\hline
100 & 0.5	&DPMLIV  		& 0.3        & 11.4	          & 100	\\
	&	& NormalLIV		&1.2 	& 9.1	&	100	\\
	& 	&	2SLS  		&  1.9	&  31.6	&	100\\
\hline
500 &	& DPMLIV	 	& 0.6  	&3.1 	&	100\\
  	& &NormalLIV 	& 0.6	& 2.8	&    100 \\
	&  & 2SLS  		&3.1 	& 12.5	&	100\\
\hline
2000   &	&  DPMLIV  			&0.1	& 1.2	& 100	\\
 	&	& NormalLIV 		& 0.1	&  1.4 	&    100\\
	& 	& 2SLS  			&0.9 	& 5	&	100\\
\hline \hline
100		&1.5 & DPMLIV	& 0.5 	& 6.3	& 100	\\
	& & NormalLIV		& 1.0	& 	6.5	& 100	\\
	& &2SLS  			&     1.7  & 9.8		& 100\\
\hline
500   	&	& DPMLIV  		&0.7      &     2.1	& 100	\\
	&	& NormalLIV 		&0.8	          &  2.5 	&  100  \\
	&	&2SLS  	 	&    2.3      & 4.2  &  18	\\
\hline
2000   &	& DPMLIV  				&0.2	  &0.7	& 	100\\
	&	& NormalLIV 				& 0.2	&  0.9	&  100 \\
	&	& 2SLS  				& 0.2	&1.6    &	100 \\
\hline
\end{tabular}
\caption{{\bf Simulation result under mixture normal outcome errors: ATE.} Mean absolute bias, mean width of posterior 95\% credible interval, and coverage of 95\% credible interval of estimated ATE under weak and strong instruments and varying sample sizes for three different methods, with $\alpha_1=0.1$, $\alpha_0=0.1$, $\alpha_D=$0.2; the errors $\epsilon_1$ and $\epsilon_0$ are generated from a 4-component mixture of Normal distributions with weights $w = (0.15,0.4,0.25,0.05)$, mean parameter vector $\mu = (-0.1,1,1,10)$, and variance vector $v= (0.01,0.1,0.1,10)$; true ATE is 15.}
\label{tab::ATE}
\end{table}

\begin{table}[!ht]
\centering
\begin{tabular}{cccccc}
\hline
 &  & & {\bf Absolute} & \multicolumn{2}{c}{\bf 95\% Credible Interval}  \\ \cline{5-6}
\bf n  & $\gamma$& \bf Method &  {\bf Bias} & \bf Width & \bf \% Coverage \\
\hline
100 & 0.5	&DPMLIV  		& 0.7        & 0.6	          & 100	\\
	&	& NormalLIV		&1.1 	& 10.3	&	100	\\
	& 	&	2SLS  		&  9.1	&  100 	&	100\\
\hline
500 &	& DPMLIV	 	& 0.2  	&3.0 	&	100\\
  	& &NormalLIV 	& 0.3	& 3.4	&    100 \\
	&  & 2SLS  		&2.5 	& 12.2	&	100\\
\hline
2000&   & DPMLIV  	 	& 0.07	& 1.2	&	100\\
	&	&  NormalLIV  	&0.12  	&	1.6&    100 \\
	&	&2SLS  		&0.6 	&3.8 	&	100\\
\hline \hline
100 &1.5	&	DPMLIV 	& 0.3	& 6.2	& 	100	\\
	& & NormalLIV	& 	0.6	& 7.2 		& 	100	\\
	& &2SLS 	& 1.6	& 10 	& 	100\\
\hline
500   	&		 &DPMLIV  	 	& 0.8	&2.1	 & 98	\\
	&	& NormalLIV			&1.1 	&  3.1	&      100\\
	&	&2SLS  			& 1.4	& 3.6 	&	96\\
\hline

2000   	&	&DPMLIV  		& 0.4	&  0.7	& 	 80\\
	&	& NormalLIV  	& 	   0.5&	1.2	 &  	70   \\
	&	&2SLS  		&0.1  & 1.6 	& 	100	\\
\hline
\end{tabular}
\caption{{\bf Simulation result under mixture normal outcome errors: CATE.} Mean absolute bias, mean width of posterior 95\% credible interval, and coverage of 95\% credible interval of estimated CATE ($\textrm{CATE}|X_3=1$) under weak and strong instruments and varying sample sizes for three different methods, with $\alpha_1=0.1$, $\alpha_0=0.1$, $\alpha_D=0.2$; the errors $\epsilon_1$ and $\epsilon_0$ are generated from a 4-component mixture of Normal distributions with weights $w = (0.15,0.4,0.25,0.05)$, mean parameter vector $\mu = (-0.1,1,1,10)$, and variance vector $v = (0.01,0.1,0.1,10)$; true CATE is 20.}
\label{tab::CATE}
\end{table}

\begin{figure}[!ht]
\centering
\includegraphics[height=5in,width=6in]{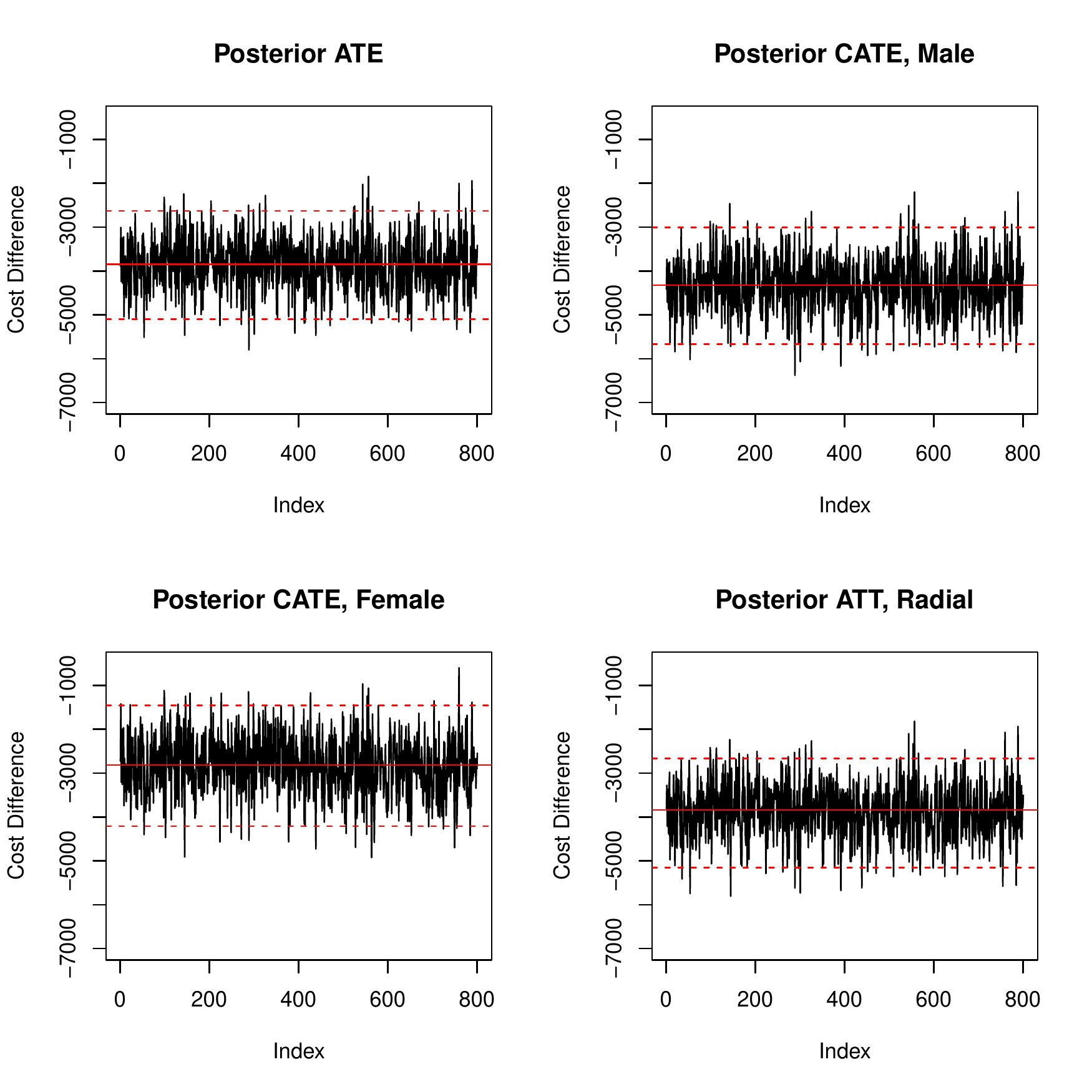}
\caption{Posterior chain of different treatment effects considered in the study. Posterior mean is represented by the horizontal solid red line, where as the 95\% credible interval region is represented by the dotted red lines. }
\label{fig::TEff}
\end{figure}

\begin{table}[!ht]
\centering
\begin{tabular}{cccccc}
\hline
 {\bf Causal}            &        &             & {\bf Absolute} & \multicolumn{2}{c}{\bf 95\% Credible Interval}  \\ \cline{5-6}
{\bf Parameter} & \bf $s^2$  & \bf Method &  {\bf Bias} & \bf Width & \bf \% Coverage \\
\hline

{\textbf{ATE}}(true = 14) 		&	 $ 0.25  $  	&  DPMLIV  	&   0.6     & 2.9 	& 95	\\
 		&	& NormalLIV 		& 0.8  	&  13.3 	& 100   \\
		& 	& 2SLS  		& 2.2 	&10.1    &88	\\
		\cline{2-6}
 		& $ 1 $  	&  DPMLIV  	&  0.6       & 2.8  	& 93	\\
 		&	& NormalLIV 		&  0.8	&  13.2	&    100 \\
		& 	& 2SLS  		& 1.8	&   10.5 & 98	\\
										\cline{2-6}
 		 & $  10 $   	&  DPMLIV  	&      0.6 &  2.6   	&    94  \\
 		&	&   NormalLIV 		&   0.7 	&   14.2 	&      100 \\
		& 	&2SLS  		&   2.4	&   11.1 &    100	\\			

								\cline{2-6}
 		 & $ 100 $   	&   DPMLIV  	&   1.4       &   2.3   	&   34\\
 		&	&   NormalLIV 		&   1.4 	&   17.16 	&   100\\
		& 	&  2SLS  		&   2.8	&   14.1   &   96 	\\			
\hline
{\textbf{CATE}}($X_3=1$) & $ 0.25 $  		& DPMLIV	& 0.8 	&  3.4	&   90   \\
(true = 18) 		& & NormalLIV			&  1.2 & 15  	& 100	\\
		& &2SLS  			&       2.6  & 14		& 97 \\
				\cline{2-6}
 		& $1$   	&  DPMLIV  	&  0.6       & 3.3  	& 92	\\
 		&	& NormalLIV 		&  0.9	&  14.5	&   100 \\
		& 	& 2SLS  		&  2.7	&   14.5 &98	\\
						\cline{2-6}
 		&  $10$   	&  DPMLIV  	&     0.8   &  3.3 	&   94   \\
 		&	&  NormalLIV 		&  1.0 	&   15.0	&    100\\
		& 	&   2SLS  		&    2.9	&     15.2 &    100	\\

						\cline{2-6}
 		& $100$   	&  DPMLIV  	&    1.3      &  2.9  	&   60\\
 		&	& NormalLIV 		&  1.3 	& 17.9 	&    100 \\
		& 	&   2SLS  		&   4.8	&   22.7   &  96	\\

\hline
\end{tabular}
\caption{{\bf {Simulation result under Gamma errors and varying variance ($s^2)$ on $\Theta$.} }Mean absolute bias, mean width of posterior 95\% credible interval and coverage of 95\% credible interval of estimated average treatment effect (ATE) and conditional average treatment effect (CATE($X_3=1$)) under Gamma(3,0.1) errors, $n= 2000$ and $\Theta \sim \text{Normal}(\mu, s^2)$. True value of $\beta_0^0$ is fixed at 90, $\beta_0$ is fixed at  $\{-0.5,1.5,0\}$, $\beta_1^0$ is fixed at 100, $\beta_1$ is fixed at $\{-0.5,1.5,10\}$, $\beta_D$ is fixed at $\{0,0,1\}$. The errors $\epsilon_1$ and $\epsilon_0$ are generated from Gamma(3,0.1) distributions; $\gamma = 1.5$; $\alpha_0 = 0.1$, $\alpha_1 = 0.1$ and $\alpha_D = 0.2$.} 
\label{tab::NPDMtheta}
\end{table}

\begin{table}[!ht]
\centering
\begin{tabular}{cccccc}
\hline
 {\bf Causal}            &        &             & {\bf Absolute} & \multicolumn{2}{c}{\bf 95\% Credible Interval}  \\ \cline{5-6}
{\bf Parameter} & \bf  $(a,b)$ & \bf Method &  {\bf Bias} & \bf Width & \bf \% Coverage \\
\hline
{\textbf{ATE}} & (0.1,1) & DPMLIV  & 0.6 	& 2.98 &   98	\\
(true = 14) 		& (10,1) 	&	DPMLIV  & 0.6  	& 2.5 	&92	 \\
\hline
{\textbf{CATE}}($X_3=1$) &(0.1,1)	& DPMLIV & 0.6 	& 3.3 & 94     \\
(true = 18) 		& (10,1) & DPMLIV 			&     0.7    & 2.8 & 92 \\
\hline
\end{tabular}
\caption{{\bf Sensitivity analysis with varying hyper parameters $(a,b)$ on the Gamma prior of the concentration parameter $c$ for $n = 2000$ and strong IV:} Mean absolute bias, mean width of posterior 95\% credible interval and coverage of 95\% credible interval of estimated average treatment effect (ATE) and conditional average treatment effect (CATE($X_3=1$)) under Gamma(3,0.1) errors. True value of $\beta_0^0$ is fixed at 90, $\beta_0$ is fixed at  $\{-0.5,1.5,0\}$, $\beta_1^0$ is fixed at 100, $\beta_1$ is fixed at $\{-0.5,1.5,10\}$, $\beta_D$ is fixed at $\{0,0,1\}$, and $\Theta$ is generated from Normal(0,0.5) distribution. The errors $\epsilon_1$ and $\epsilon_0$ are generated from Gamma(3,0.1) distributions; $\gamma = 1.5$; $\alpha_0 = 0.1$, $\alpha_1 = 0.1$ and $\alpha_D = 0.2$. The hyper parameters are presented in the increasing order of prior mean for the concentration parameter $c$. Other prior specifications remain unchanged.}
\label{tab::ATE}
\end{table}

\begin{table}[!ht]
\centering
\begin{tabular}{cccccc}
\hline
 {\bf Causal}            &        &             & {\bf Absolute} & \multicolumn{2}{c}{\bf 95\% Credible Interval}  \\ \cline{5-6}
{\bf Parameter} & \bf  $(\Psi^{-1},\nu)$ & \bf Method &  {\bf Bias} & \bf Width & \bf \% Coverage \\
\hline
{\textbf{ATE}}
	& (1,4)	& DPMLIV &0.9		& 2.0 	& 90  	\\
(true = 14) 		& (0.5,4) 	&	DPMLIV  &  0.9  	& 3.1 	& 98	 \\
\hline
{\textbf{CATE}}($X_3=1$) 
		&  (1,4) &DPMLIV	& 1.1   & 3.5 	& 85	\\
(true = 18) 		& (0.5,4) & DPMLIV 			&     1.1    & 3.7 & 85 \\
\hline
\end{tabular}
\caption{{\bf Sensitivity analysis with varying hyper parameters $(\Psi^{-1},\nu)$ on the Inverse-Wishart prior for the variance of base distribution $G_0$, with $n = 2000$ and strong IV:} Mean absolute bias, mean width of posterior 95\% credible interval and coverage of 95\% credible interval of estimated average treatment effect (ATE) and conditional average treatment effect (CATE($X_3=1$)) under Gamma(3,0.1) errors. True value of $\beta_0^0$ is fixed at 90, $\beta_0$ is fixed at  $\{-0.5,1.5,0\}$, $\beta_1^0$ is fixed at 100, $\beta_1$ is fixed at $\{-0.5,1.5,10\}$, $\beta_D$ is fixed at $\{0,0,1\}$, and $\Theta$ is generated from Normal(0,0.5) distribution. The errors $\epsilon_1$ and $\epsilon_0$ are generated from Gamma(3,0.1) distributions; $\gamma = 1.5$; $\alpha_0 = 0.1$, $\alpha_1 = 0.1$ and $\alpha_D = 0.2$. Hyper-parameters are presented in the increasing order of informativeness, less informative to highly informative prior variance. Other prior specifications remain unchanged.}

\label{tab::ATE}
\end{table}

\end{document}